\documentstyle[preprint,aps,psfig,prl]{revtex}

\def \ef     {E_{\rm F}}
\def \ee     {E_{e\mbox{-}e}}
\def \eeint  {{e\mbox{-}e}}
\def \mup    {\mu_{\rm peak}}
\def \mcrit  {\mbox{{\bf C}}^*}
\def \mm     {\mbox{{\bf M}}}

\begin{document}

\title{Dilute electron gas near the metal-insulator transition
in two dimensions.}

\author{Alexander Punnoose and Alexander M. Finkel'stein}

\address{Department of Condensed Matter Physics, 
Weizmann Institute of Science,\\
Rehovot, 76100, Israel}

\maketitle

%
%
\newpage

{\bf 
In recent years systematic experimental studies of the 
temperature dependence of the resistivity in a variety
of dilute, ultra clean two dimensional electron/hole
systems have revived the fundamental question of localization
or, alternatively, the existence of a metal-insulator transition
in the presence of strong electron-electron interactions in two dimensions.
We argue that under the extreme conditions of
ultra clean systems not only is the electron-electron interaction 
very strong but the role of other system specific properties are also 
enhanced. In particular, we emphasize the role of valleys in determining 
the transport properties of the dilute electron gas in silicon inversion
layers (Si-MOSFETs). 
It is shown that for a high quality sample
the temperature behavior of the resistivity
in the region close to the critical region of the 
metal-insulator transition is well described 
by a renormalization group analysis of 
the interplay of interaction and disorder
if the electron band is assumed to have two distinct valleys.
The decrease in the resistivity up to five times 
has been captured in the correct temperature
interval by this analysis, without involving any 
adjustable parameters. 
The considerable variance in the data obtained
from different Si-MOSFET samples is attributed to the
sample dependent scattering rate across the two valleys,
presenting thereby with
a possible explanation for the absence of universal
behavior in  Si-MOSFET samples of different quality.
}


The resistivity in a variety of high mobility
two dimensional (2D) electron/hole systems is
seen experimentally to exhibit a number of interesting
anomalies that do not, as yet, have an adequate theoretical 
understanding. 
(For an extensive bibliography, see Ref.~\ref{cite:reviews}.)
The high quality of the samples allows  measurements
to be made at very low carrier densities
corresponding to $r_{s}\gtrsim 10$, where $r_{s}=\ee/\ef$ 
is the ratio of the Coulomb energy to Fermi energy.
When the resistivity at high temperatures is 
comparable to or less than the quantum resistance, $h/e^2$, 
the resistivity, $\rho(T)$,  drops noticeably as the
temperature is reduced \cite{cite:rhoT1,cite:rhoT2}.
The drop appears to be completely quenched when a magnetic 
field is applied parallel to the plane 
\cite{cite:rhoB1,cite:rhoB2,cite:rhoB3}. 
This anomalously strong 
positive magnetoresistance, which is obviously related to 
the spin degrees of freedom, points to the importance of the 
electron-electron ($\eeint$) interaction in this phenomenon.
%

These experimental observations have revived interest in 
the fundamental question of the existence of a
metal-insulator transition in 2D systems in the presence of a 
strong $\eeint$ interaction.
(We use the term ``transition'' to describe the 
qualitative change in the resistivity at the experimentally accessible 
temperatures, not discussing the question of the limit $T\rightarrow 0$.)

Although the drop in the resistivity is seen in almost all the
different dilute systems studied so far, and are therefore 
generally considered to be universal, quantitative comparison 
indicates that the magnitude of the effect is very sensitive to 
the system used. The most pronounced anomaly has been reported 
in the cleanest (001) Si-MOSFET samples, where a steep drop in 
$\rho (T)$ of up to five to six times has been 
observed.

In Fig.~\ref{fig:Si15} the temperature dependence of the 
resistivity for different densities in a Si-MOSFET sample
with a very high peak mobility of 
$\mup=$~32,000~[cm$^{2}$/Vs] (measured at $T=0.3$K)
has been reproduced \cite{cite:fig1}. The insulating region,
where the resistivity increases with decreasing temperature,
is labeled as {\bf I} in Fig.~\ref{fig:Si15}.  
The range of densities where the resistivity depends non-monotonically 
on temperature is labeled as $\mcrit$ in Fig.~\ref{fig:Si15}.
(For the specific sample used in  Fig.~\ref{fig:Si15}, 
this region covers electron densities in the range 
$0.8\times10^{11}<n<1\times10^{11}$cm$^{-2}$.)
A narrow range of densities in between these two regions,
in which the separatrix that separates the insulating phase 
from the metallic phase should lie (if a true metal-insulator 
transition exists) can
be considered as the critical region {\bf C}. 
The region where no clear maximum
in the resistivity is observed (unlike $\mcrit$) is labeled as $\mm$ in 
Fig.~\ref{fig:Si15}. 
The region $\mcrit$ is the subject of interest of this paper.

Since the maximum of the resistivity in the
region $\mcrit$ is comparable to $\sim h/e^{2}$, 
the transport mean free path time, $\tau $, of the electrons 
here is such that $\hbar /\tau \lesssim \ef$. The maximum
together with the steep decrease in $\rho (T)$ occurs at low 
temperatures well below the Fermi energy 
(see Fig.~\ref{fig:Si15}). Therefore, the indications are that 
the non-monotonic behavior of the resistivity in the
region $\mcrit$ is a manifestation of the physics of 
strongly interacting electrons that are in the diffusive regime: 
$T<\hbar /\tau \lesssim \ef$.

Away from the region $\mcrit$ and deep in the region labeled 
$\mm$ in Fig.~\ref{fig:Si15} a naive estimate for $\tau $ can 
be extracted from the Drude expression for the resistivity, 
$\hbar /\tau =4(e^{2}/h)\rho (T=0)\ef$, 
with $\rho (T=0)$ being the extrapolated value of the 
resistivity at $T=0$. 
This estimate gives values for $\hbar /\tau $ that are well 
below the Fermi energy, while the steep drop in the resistivity 
develops at temperatures that are comparable to or larger 
than $\hbar /\tau $. This implies that the anomalies
in the region $\mm$ occur in the temperature range
$\hbar /\tau \lesssim T<\ef$, and their origin may 
be attributed (at least partially) to a strong 
temperature dependence of the single particle mean free path time 
$\tau (T)$ \cite{cite:stern,cite:gold,cite:SDSarma}. 
On the contrary, the situation in the region 
$\mcrit$ is quite different since the anomalies here are in the 
temperature range $T<\hbar/\tau \lesssim \ef$ where 
disorder quenches the effects of thermal smearing 
on $\tau (T)$ \cite{cite:SDSarma}.

These considerations lead us to the conclusion that the anomalous
decrease in the resistivity in the two regions, 
$\mcrit$ and $\mm$, may have different origins and are hence best 
studied separately. In this paper we analyze
the transport properties in the region $\mcrit$, close to the
critical region {\bf C}, 
where the transport is controlled by the propagation of diffusive
collective modes. We demonstrate that the phenomenon in this 
region can be understood within the framework of a theory 
describing the effect of the $\eeint$ interaction on 
the propagation of these modes \cite{cite:aa}. The peculiarity 
of dilute conductors is that at low temperatures the 
antilocalizing component of this effect becomes dominant.

Although universal behavior is generally expected to hold in the
critical region, no universal scaling relating the $\rho (T)$ 
curves from different samples has been found. 
In fact, considerable variance is seen even in the data obtained
from different Si-MOSFET samples of similar origin,
including the data from the same sample that varies
due to the degradation of the high mobility sample with 
time.
Hence, for a quantitative understanding of the
temperature dependence of the resistivity in the 
region not far from the transition some system specific 
non-universal mechanism should be necessarily invoked. 
The conduction band of the electrons 
in a (001) Si-MOSFET surface has two almost degenerate valleys 
located at points $\pm Q_{0}$ ($Q_{0}=0.85\times(2\pi/a)$, where
$a$ is the lattice constant of Si) \cite{cite:ando}.
In what follows, the sensitivity of the transport properties 
of the dilute electron gas to the scattering rate across the 
two valleys is presented as a possible explanation for the 
absence of this universality.  
At temperatures comparable to the rate of the intervalley 
scattering, $\hbar /\tau_{\perp }$, a crossover occurs 
between a band with
two distinct valleys and a band where the two valleys are 
effectively unified due to the intervalley scattering. 
We believe that in a typical sample the value of 
$\hbar/\tau_{\perp }$ falls within the temperature
interval in which measurements are made. Hence
due to the crossover at $T\sim\hbar/\tau_{\perp}$ the resistivity
$\rho(T)$ will be non-universal.
Only in an ultra clean sample (like the one presented in 
Fig.~\ref{fig:Si15}), where the intervalley scattering is very 
weak and the two valleys are well separated, should a universal 
behavior hold.

To understand the temperature dependence of $\rho (T)$ in the 
case of two valleys, we study the interplay of the appropriate 
collective modes.
These modes describe fluctuations of the local density of 
particles, spin, and in addition the fluctuations involving 
electron states from different valleys. 
The evolution of the collective modes, in the limit of long
wavelengths and small frequencies, are described by a 
singular propagator with a diffusion pole 
$\propto 1/(Dq^{2}-i\omega )$, 
where $D$ is the diffusion constant.
(Note that the collective modes may exist even when a
description in terms of single particle excitations is not 
possible.) These singular propagators when combined with the 
$\eeint$ interaction are known in
2D to lead to the appearance of non-analytical corrections to the
resistivity. 
On the other hand, the amplitudes of the $\eeint$ interaction
that affects the propagation of the collective modes are 
themselves known to have divergent corrections due to the 
disorder. 
The program to self-consistently take into account these 
corrections, which in fact corresponds to a derivation of a 
system of renormalization group (RG) equations, has been 
realized to lowest order in the resistivity (disorder),
and fortunately to all orders in the $\eeint$ interaction 
amplitudes \cite{cite:sashaT1}. 
The latter fact is important when 
$r_{s}$ is large
as then the electron liquid may be not far from some 
magnetic instability, implying that the Landau's Fermi
liquid amplitude $\gamma_{2}$ 
that controls the interaction of the spin-density
fluctuations may be not small.

The diffusion propagators of the electron-hole pairs in the 
presence of valleys in addition to the momentum and spin 
quantum numbers are labeled by quantum numbers 
$|\tau \rangle $, where $|\tau \rangle =\pm $ are the two
valley indices similar to the up and down spin states 
$|\sigma \rangle =\uparrow,\downarrow $. 
Altogether there are $4\times 4=16$ electron-hole
states that break up into one singlet and fifteen multiplet 
states. 
In the case of strong intervalley scattering, however, 
the modes that are made of states from different valleys 
acquire a gap proportional to $\hbar/\tau_{\perp}$. 
This implies that for temperatures, or frequencies, less than 
$\hbar /\tau_{\perp }$ such modes do not yield diverging 
contributions and hence become ineffective. 
(This is the origin of the crossover discussed above.) 
As a result, off the 16 modes only one spin singlet and 
three spin triplet combinations retain a diffusion pole. 
Therefore, the situation when the two
valleys are effectively unified becomes equivalent to the case 
with no valleys (but with the density of states doubled). 
In 2D the leading divergences are logarithmic and the RG 
equation describing the evolution of the resistivity
in the absence of valleys has been previously 
derived 
\cite{cite:sashaT1,cite:sashaT2,cite:sashaT3,cite:castellani1,cite:castellani2}: 
\begin{equation}
\frac{dg}{d\xi }=g^{2}\left[1+1-3
\left(\frac{1+\gamma _{2}}{\gamma _{2}}
\ln (1+\gamma _{2})-1\right) \right]~.
\label{eqn:1-val-g}
\end{equation}
Here $\xi=-\ln (T\tau )$ and the dimensionless parameter
$g=(e^{2}/\pi h)\rho$; 
note that apart from the standard unit $e^{2}/h$ an additional
factor $1/\pi $ has been introduced in $g$. 
In the square brackets the first term corresponds to the weak 
localization correction (quantum interference) \cite{cite:g4},
while the second term is the contribution of the singlet 
density mode which due to the long ranged nature of the Coulomb 
interaction is also universal \cite{cite:aa}.
The last term describes the contribution of the three triplet 
modes. Due to the difference in symmetry of the singlet and 
the triplet wave functions under exchange the last two terms 
have opposite signs that favor localization
and antilocalization, respectively. 
The resulting flow of $g(\xi )$ becomes antilocalizing when 
$\gamma_{2}$ is greater than the value $\gamma _{2}^{\ast}=2.04$. 
This value demands, however, the presence of rather strong 
electron correlations. 
(For comparison, the effective amplitude of the $\eeint$
interaction for small momentum transfer with $n$ valleys participating
in screening equals $1/2n$; the suppression occurs due to the increased
screening.)

In the case of two distinct valleys, {\em i.e.}, when 
$T>\hbar /\tau_{\perp}$, Eq.~\ref{eqn:1-val-g} can be easily 
generalized as: 
\begin{equation}
\frac{dg}{d\xi }=g^{2}\left[2+1-15\left( 
\frac{1+\gamma_{2}}{\gamma_{2}}
\ln (1+\gamma _{2})-1\right) \right]~.
\label{eqn:2-val-g}
\end{equation}
The difference between the numerical factors in 
Eq.~\ref{eqn:1-val-g} and Eq.~\ref{eqn:2-val-g} are easily 
accounted for in terms of the number of
degrees of freedom in each case. 
Naturally, the weak localization term
becomes twice as large because of the presence of two valleys. 
Although the $\eeint$ amplitude controlling the interaction in 
the singlet channel is reduced by a factor of two as electrons 
from both valleys participate in screening, after adding together 
the contributions from each valley  the second term in 
Eq.~\ref{eqn:2-val-g} remains the same as that in 
Eq.~\ref{eqn:1-val-g} \cite{cite:aa}. 
Finally, the difference in the number of the multiplet modes
increases the coefficient of the $\gamma _{2}$ term from $3$ 
to $15$. (Now, $\gamma _{2}$ is the Fermi liquid amplitude that 
controls the $\eeint$ interaction in all the multiplet channels. 
Like in Eq.~\ref{eqn:1-val-g} this dimensionless parameter is 
normalized by the density of states for a single spin and valley 
species.) 
As a result of these modifications the value of $\gamma_{2}$ 
required for the flow of $g(\xi )$ to become antilocalizing is
considerably reduced to $\gamma _{2}^{\ast }=0.45.$ 
(This value is not too far from the $\eeint$ interaction amplitude for 
small angle scattering, that we use to compare, which equals 
0.25 for $n=2$.)
The reduction 
of $\gamma _{2}^{\ast }$ from $2.04$ to $0.45$ makes it easier 
in the case of two valleys to reach the stage where the 
resistivity starts to decrease.

In conventional conductors the initial values of the amplitude 
$\gamma_{2}$ are small, and the net effect is in favor of 
localization. 
In dilute systems, however, this amplitude is enlarged due to 
$\eeint$ correlations. In addition, in 2D the amplitude 
$\gamma _{2}$ also experiences logarithmic
corrections due to the 
disorder 
\cite{cite:sashaT1,cite:sashaT2,cite:sashaT3,cite:castellani1,cite:castellani2}. 
The equation describing the 
RG evolution of $\gamma _{2}$ is the same for both one and 
two valleys: 
\begin{equation}
\frac{d\gamma _{2}}{d\xi }=g~\frac{(1+\gamma _{2})^{2}}{2}~.
\label{eqn:val-r}
\end{equation}
It follows from this equation that as the temperature is 
lowered $\gamma _{2}$ increases monotonically. When it
increases beyond the value $\gamma _{2}^{\ast}$ the 
resistivity will pass through a maximum. 
Although the initial values of $g$ and $\gamma_{2}$ are not 
universal and depend on the system, the flow of 
$g$ according to the RG equations can be described by a
universal function $R(\eta )$ \cite{cite:sashaT1}:
\begin{equation}
g=g_{\max}R(\eta)\text{ and }\eta=g_{\max}\ln(T_{\max }/T)~,
\label{eqn:rgsol}
\end{equation}
where $T_{\max}$ is the temperature at which $g$ reaches 
its maximum value $g_{\max }$, {\em i.e.}, 
$\gamma_{2}(T_{\max })=\gamma _{2}^{\ast }$. 
For the case of two valleys, the function $R(\eta )$ is 
found here by numerically  integrating 
Eq.~\ref{eqn:2-val-g} and Eq.~\ref{eqn:val-r} 
with the boundary conditions: 
$g(\xi=0)=g_{\max }=1$ and 
$\gamma _{2}(\xi =0)=\gamma _{2}^{\ast }=0.45$.

Thus, if the experimental data of the resistivity are 
scaled with respect to the maximum value as $g/g_{\max }$ 
and plotted against $\eta $, which involves scaling the 
log of the temperature by $g_{\max }$, then the
data should collapse on the function $R(\eta )$. 
This analysis has certain limitations, however. 
The RG equations have been derived in the lowest order
in $g$ and therefore cannot be applied in the critical
region {\bf C}  where $g\gtrsim 1$. 
On the other hand, for $g\ll 1$ exponentially small temperatures 
are needed for changes in the resistivity to become noticeable. 
In addition, some other (not yet completely identified) 
mechanism operating in the region $\mm$ may mask the
discussed logarithmic corrections that are very weak when $g\ll 1$.

For these reasons, only curves in the region $\mcrit$ with 
maximum $g$ ranging from $g_{\max }\approx 0.3$ to 
$g_{\max }\approx 0.6$ have been used to test the RG analysis. 
The result is presented in Fig.~\ref{fig:rgsol}, where the 
data  has been scaled as in Eq.~\ref{eqn:rgsol}. 
The decrease in the resistivity up to five times together with
its saturation has been captured in the correct temperature 
interval by this analysis. Note that no adjustable parameters 
were used in the procedure.

We emphasize again that this universal behavior will 
be observed only in ultra clean samples, and will 
not be found in samples that
are only moderately clean, because of the
crossover at $T\sim \hbar/\tau_\perp$. Next, in samples
with a low mobility, where a description in terms
of an effective single valley is relevant, the 
large value for $\gamma_2^\star=2.04$ makes
it difficult for the non-monotonicity to be observed
as the initial values of $\gamma_2$ are, most probably, far
away from $2.04$. Then, to scale the amplitude $\gamma_{2}$ 
till the value $\gamma _{2}^{\ast}$ will, for $g\ll 1$, 
demand exponentially small temperatures as the corrections 
depend on the temperature only logarithmically. 
On the other hand, for $g$ near the critical region,
where changes in the resistivity develops rapidly, 
the resistivity flows to
such large values that the system instead of passing through 
the maximum becomes insulating.
To summarize, we have argued that it is not the large value of
$r_{s}$ that makes the physics 
of the region not far from the transition
in high mobility MOSFET samples 
so different from that in lower mobility samples, but 
the difference in their number of effective valleys. 
Note that in some samples the discussed 
anomalies have not been observed even 
for $r_{s}\approx 10$.

The strong magnetoresistance in a parallel magnetic field can be 
also understood by the reduction of the number of diffusion 
modes that contribute to the antilocalizing 
corrections \cite{cite:tvrB,cite:castellaniB}. Here, 
the Zeeman splitting induces a gap in the propagators of the 
diffusion modes that are made of states with
different spin projections. As a result these modes will no
longer contribute to the antilocalization corrections. In a very 
strong magnetic field when the electrons are 
completely polarized, the system becomes identical to one with
no valleys with the original valleys acting as 
fictitious spin projections. 
The difference in the resistivity of two- and 
one-valley systems, which is large at low enough temperatures, 
will be recovered as the magnetic field is applied resulting in 
a very strong positive magnetoresistance.

{\em In conclusion}, we have demonstrated that in an 
ultra clean (100) Si-MOSFET the temperature behavior of the 
resistivity in the region $\mcrit$ is well described by the 
RG analysis of the interplay of the $\eeint$ 
interaction and disorder when the electron band has two 
distinct valleys. For $g$ not too large, the system of RG 
equations in the case of two valleys is an internally 
consistent theory (for all practical purposes), unlike 
that for a single valley where $\gamma _{2}$ diverges at 
$\eta \approx 1$ after the maximum of $g$ is passed. 
This divergence points to some instability of a magnetic 
nature in the electron gas, beyond which the RG analysis in 
its present form does not hold anymore. This instability 
also occurs in the case of two valleys but at such low 
temperatures that it has no practical significance.

Finally, a few remarks concerning the electron gas in Si-MOSFETs.

The intervalley scattering involves a transfer of a large 
momentum $2Q_{0}$ in the direction perpendicular to the 
conduction plane. The width of the extension $z_{0}$
of the electrons inside the inversion layer is larger than 
the atomic scale $a$, with $z_{0}$ becoming even larger when 
the electron density decreases. Therefore, the calculation 
of the intervalley scattering amplitudes involves an integration 
of smooth electron wave functions together with a fast 
oscillating factor with $2Q_{0}$ Fourier component.
As a result one gets amplitudes proportional to a high power 
of the parameter $1/(Q_{0}z_{0})$, which is small. Hence, the 
amplitudes of this process involving Coulomb interactions are 
very weak. The imperfections on the interface, on the other hand, 
can be of the atomic scale and their matrix element will contain 
Fourier components of high momenta. We assume, therefore, that 
the rate of the intervalley scattering is controlled by
the quality of the interface, which is sample dependent. 

Some information about the rate of the intervalley scattering 
can be obtained from the magnetoresistance measurements in a 
weak magnetic field perpendicular to the conduction plane. 
The results of these measurements \cite{cite:digamma}, which yield a negative 
magnetoresistance, have been fitted with a standard expression
containing the Digamma function, $\Psi$. 
Depending on the rate 
of the intervalley scattering, the theory predicts different 
values for the prefactor $\alpha$ in this expression: 
$\alpha=1$ in the absence of the intervalley scattering and 
$\alpha=0.5$ in the case of strong intervalley
scattering \cite{cite:fukuyama}. 
The experimental situation for the sample used in
Fig.~\ref{fig:Si15} (but after some age degradation, however)
remains uncertain. The optimal fit gives values for $\alpha$ 
between $0.6$ and $0.8$, with a tendency to be larger when the 
density decreases \cite{cite:WL}. 
We consider the fact that $\alpha$ is 
noticeably larger than $0.5$ as an indication that the 
intervalley scattering is not too strong in the system
at low density. 

In our analysis throughout we have ignored the valley splitting. 
In the absence of intervalley scattering the valley splitting 
does not influence the magnetoresistance. However, the combined 
effect of the intervalley scattering and the valley splitting 
will suppress the coefficient $\alpha$ below $0.5$. 
Since the actual coefficient is larger than $0.5$, we have
ignored the valley splitting. It is also known from 
theoretical calculations that the valley splitting is small at 
low densities that are of interest to us here \cite{cite:ando}. 

The chiral splitting 
of the electron band due to the spin orbit interaction in the presence
of the asymmetric interface potential has been often discussed in connection
with the dilute electron gas. This
mechanism is, however, 
incompatible~\cite{cite:SOtheory1,cite:SOtheory2,cite:SOtheory3}
with the observed negative magnetoresistance in MOSFETs \cite{cite:WL}.
Since actually there are no reasons to expect a considerable chiral 
splitting in n-type semiconductors, we have ignored the 
spin orbit effects in our analysis of the dilute electron 
gas in Si-MOSFETs.


\newpage

\acknowledgements
We have benefited greatly from numerous discussions with
V. M. Pudalov and we appreciate his contribution
to our understanding of the experimental data.
This work was supported by the Israel Science 
Foundation - Centers of Excellence Program, and by the
German Ministry of Science (DIP).
A. P. acknowledges the support of the Feinberg Graduate School
of the Weizmann Institute of Science.


\begin{figure}
\caption{Resistivity of a high mobility Si-MOSFET sample 
for various densities as a function of temperature.
The electron densities, $n$, are defined
in units of $10^{11}$cm$^{-2}$. Data are reproduced from
Fig. 1(a) of Ref.~\protect\ref{cite:fig1}. {\bf I} labels
the insulating region, and {\bf C} labels the critical region. 
$\mcrit$ is the region near the critical region where a 
clear maximum is observed in the temperature dependence of
the resistivity and is the region that is studied in the text.
$\mm$ labels the region further away from the critical region where such a 
maximum is not observed.
}
\label{fig:Si15}
\end{figure}

\begin{figure}
\caption{The data corresponding to $n=0.83, 0.88$, and $0.94$
$\times 10^{11}$cm$^{-2}$ in Fig.~\protect\ref{fig:Si15} are scaled 
according to Eq.~\protect\ref{eqn:rgsol}. The solid line corresponds
to the solution of the renormalization group Eqs.~\protect\ref{eqn:2-val-g}
and \protect\ref{eqn:val-r}; no adjustable parameters
have been used in this fit.
}
\label{fig:rgsol}
\end{figure}

\newpage

\pagestyle{empty}
\centerline{\psfig{figure=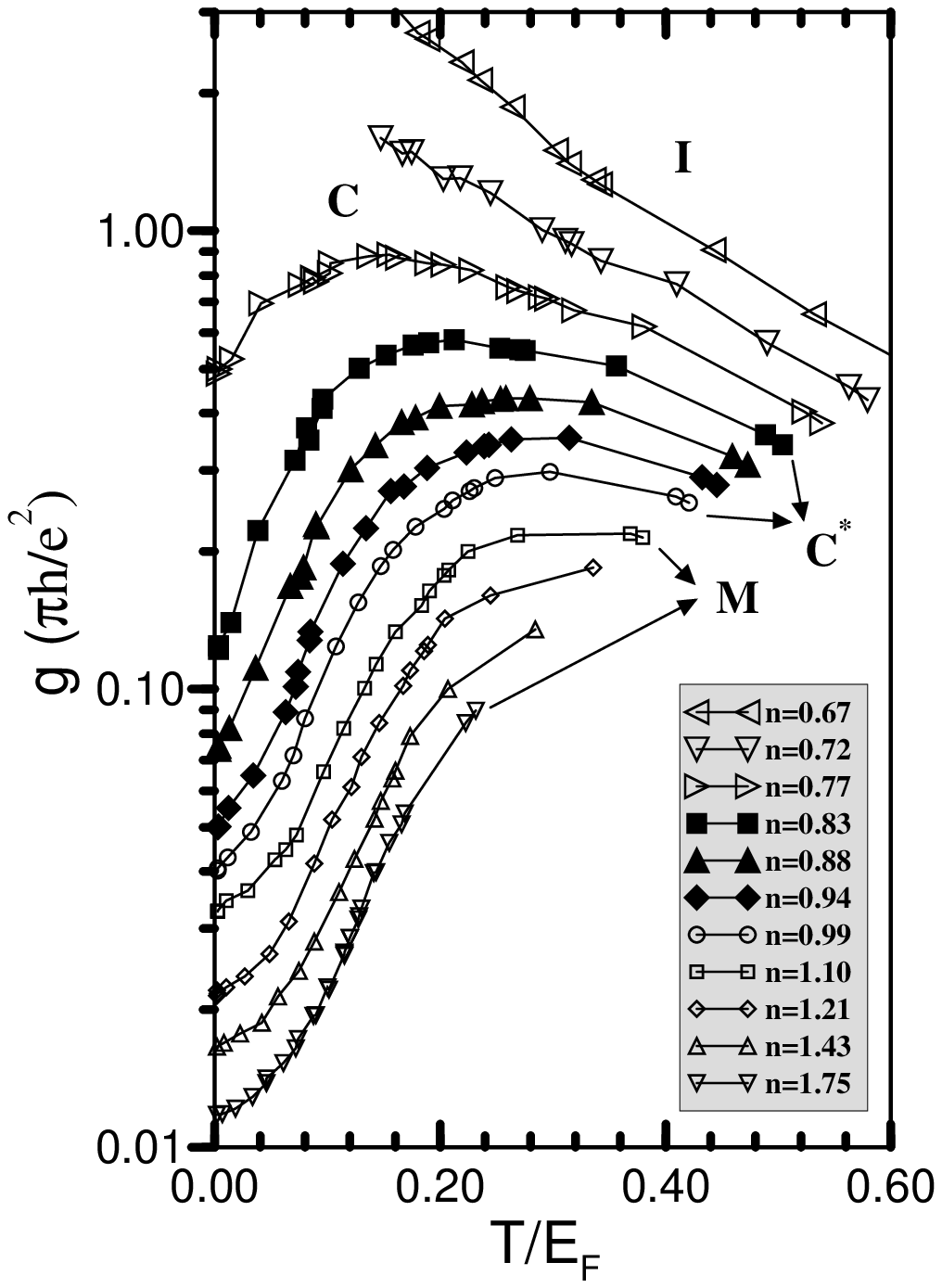}}
\vspace{0.25\textheight}
{\centerline{Figure 1., A. Punnoose, A. M. Finkel'stein}}

\newpage
\centerline{\psfig{figure=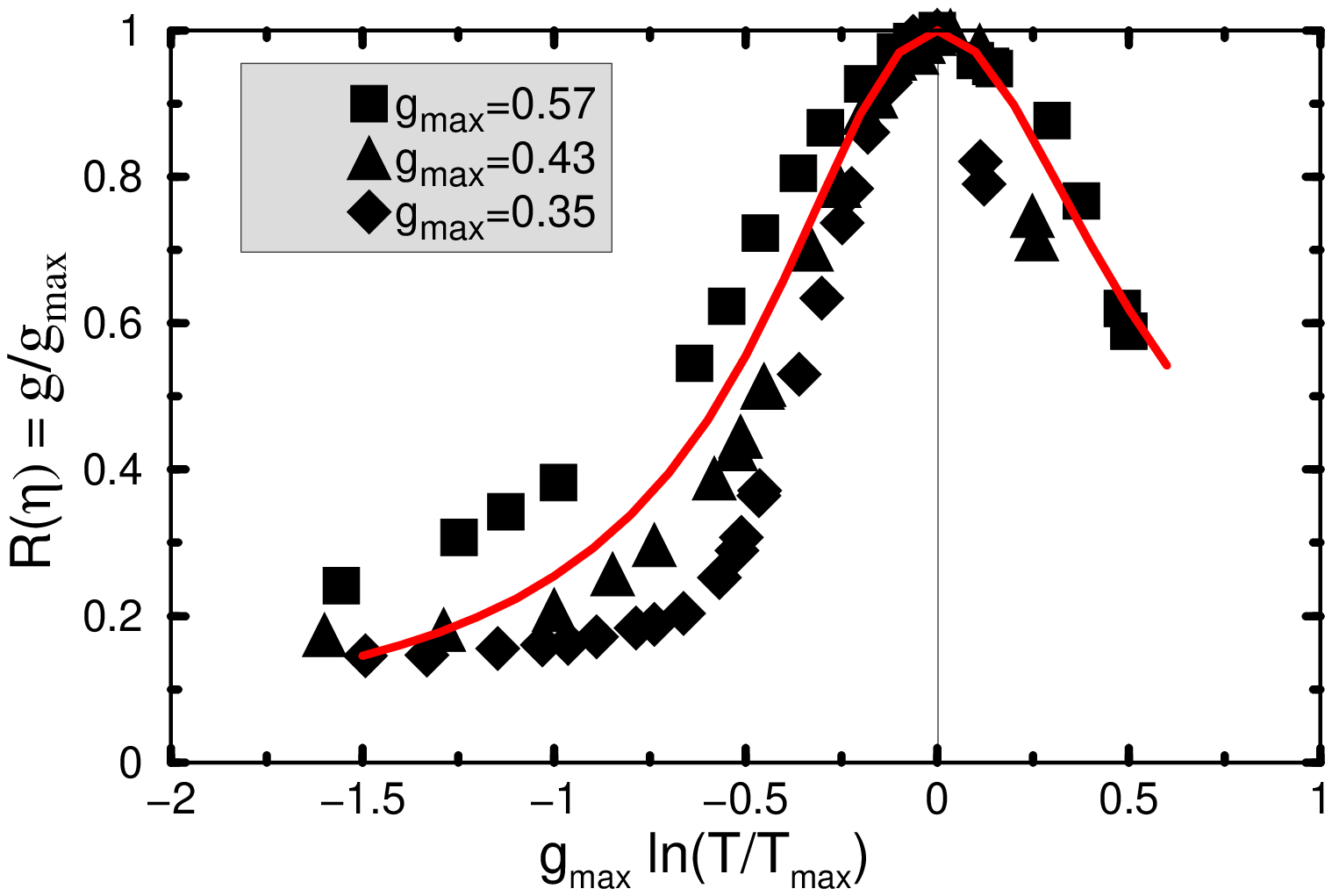}}
\vspace{0.45\textheight}
{\centerline{Figure 2., A. Punnoose, A. M. Finkel'stein}}

\end{document}